\documentclass[12pt]{iopart}

\usepackage{latexsym}
\usepackage{eucal}
\usepackage[latin1]{inputenc}
\usepackage[english]{babel}
\usepackage{wasysym}
\usepackage{amsfonts}
\usepackage{amssymb}
\usepackage[dvips]{graphicx}

\newcommand{\ba}{\begin{align}}
\newcommand{\ena}{\end{align}}
\newcommand{\be}{\begin{equation}}
\newcommand{\ee}{\end{equation}}
\newcommand{\fr}{\frac}

\def\cB{{\cal B}}

\def\cG{{\cal G}}
\def\cJ{{\cal J}}

\def\cC{{\cal C}}
\def\cL{{\cal L}}

\def\cF{{\cal F}}

\def\cA{{\cal A}}

\def\cR{{\cal R}}
\def\cS{{\cal S}}

\def\cH{{\cal H}}

\def\cA{{\cal A}}

\def\dsll{\not {\! \pr}}

\def\psisl{\not {\! \! \psi}}

\def\pe{\prime}
\def\eq{\equiv}

\def\pr{\partial}
\def\btd{\bigtriangledown}

\def\pe{\prime}
\def\3s{{s \choose 3}}
\def\4s{{s \choose 4}}
\def\5s{{s \choose 5}}
\def\6s{{s \choose 6}}
\def\12{\frac{1}{2}}
\def\fr{\frac}
\def\pr{\partial}
\def\prd{\partial \cdot}
\def\btd{\bigtriangledown}

\def\a{\alpha}
\def\b{\beta}
\def\g{\gamma}
\def\G{\Gamma}
\def\d{\delta}

\def\e{\epsilon}

\def\h{\eta}

\def\L{\Lambda}
\def\m{\mu}
\def\n{\nu}

\def\r{\rho}

\def\s{\sigma}

\def\t{\tau}

\def\f{\phi}

\def\vf{\varphi}

\def\ps{\psi}

\def\cJ{{\cal J}}
\def\cC{{\cal C}}
\def\cL{{\cal L}}

\def\cF{{\cal F}}

\def\cA{{\cal A}}

\def\cR{{\cal R}}
\def\cS{{\cal S}}

\def\cH{{\cal H}}

\def\cA{{\cal A}}

\def\nn{\nonumber}

\def\be{\begin{equation}}
\def\ee{\end{equation}}
\def\bea{\begin{eqnarray}}
\def\eea{\end{eqnarray}}
\def\ba{\begin{array}}
\def\ea{\end{array}}

\def\dsll{\not {\! \pr}}
\def\psisl{\not {\! \! \psi}}

\def\e{\epsilon}

\begin{document}

\begin{flushright}
RM3-TH/06-2 \\
AEI-2006-005 \\
CERN-PH-TH/2006-009\\
hep-th/yymmddd
\end{flushright}

\title[Higher-Spin Geometry and String Theory]{Higher-Spin Geometry and String Theory}

\author{Dario Francia}

\address{ Dipartimento di Fisica, Universit\`a di Roma Tre and INFN, Sezione di Roma III\\
Via della Vasca Navale 84, I-00146 Roma \ Italy\\ and \\
Max Planck Institute for Gravitational Physics (Albert Einstein
Institute) \\ Am Mühlenberg 1 D-14476 Potsdam \ Germany
\footnote[1]{EU pre-doctoral Fellow. Address after October 1, 2006:
Institute of Fundamental Physics, Chalmers University of Technology,
412 96 Goteborg \ Sweden.} } \ead{francia@fis.uniroma3.it,\
Dario.Francia@aei.mpg.de}

\author{Augusto Sagnotti}

\address{Ph-Th Department, CERN, 1211 Geneva 23 \ Switzerland \\ and \\
Scuola Normale Superiore\footnote[2]{Permanent address after
September 1, 2006.} and INFN, Piazza dei Cavalieri 7, I-56126 Pisa \
Italy} \ead{sagnotti@sns.it}

\begin{abstract}
The theory of freely-propagating massless higher spins is usually
formulated via gauge fields and parameters subject to trace
constraints. We summarize a proposal allowing to forego them by
introducing only a pair of additional fields in the Lagrangians. In
this setting, external currents satisfy usual Noether-like
conservation laws, the field equations can be nicely related to
those emerging from Open String Field Theory in the low-tension
limit, and if the additional fields are eliminated without
reintroducing the constraints a geometric, non-local description of
the theory manifests itself.

\end{abstract}

{\it \noindent Based on the talks presented by D.F. at the ``First
Meeting of Young Researchers of the European Superstring Theory
Network'', Heraklion, Crete, October 17-28 2005 and at the ``Annual
Meeting of the Italian String Theory Network'', Pisa, November 4-5
2005, and on his Ph.D. Thesis, to be defended at Universit\`a Roma
Tre on February 2006, and on the talk presented by A.S. at the
``Fourth Meeting on Constrained Dynamics and Quantum Gravity'', Cala
Gonone (Sardinia, Italy), September 12-16 2005.}

\maketitle

\section{Introduction}\label{sec:Intro}

In four dimensions a massive spin-$s$ boson can be described via a
totally symmetric rank-$s$ tensor $\vf_{\, \m_1 \dots \m_s}$ subject
to the conditions \cite{fierz}
\begin{eqnarray} \label{fierz}
&(\Box \, + \, m^2) \, \vf_{\, \m_1 \dots \m_s} \, = \, 0 \, , \nonumber\\
&\pr^{\, \m_1} \vf_{\, \m_1 \dots \m_s} \, = \, 0 \, ,  \\
&\vf^{\, \n}_{\, \, \, \, \n \mu_1 \dots \m_{s-2}} \, = \, 0 \, ,
\nonumber
\end{eqnarray}
where the two constraints on the trace and the divergence remove the
polarizations not within a single \emph{irreducible} representation
of the Poincar\'e Group. Their distinct role was clearly recognized
long ago by Fierz and Pauli \cite{fierzpauli} when they tried to
analyze the interaction with an electromagnetic field. While the
first two equations in (\ref{fierz}) became incompatible, the trace
conditions proved harmless from the point of view of the minimal
substitution, and they proposed to proceed in the search for an
action principle leaving them as \emph{algebraic constraints}, a
choice that was to influence all subsequent research in the field.
The Fierz-Pauli program of constructing an action principle leading
to eqs.~(\ref{fierz}) was completed only in 1974 by Singh and Hagen
\cite{singhag}, who also confined their attention to \emph{massive}
fields, while the massless limit of their Lagrangians was
subsequently investigated by Fronsdal \cite{fronsdal} and Fang and
Fronsdal \cite{fang}. The resulting abelian gauge theories have been
regarded since then as a paradigm for the free dynamics of
higher-spin gauge fields.

In the Fronsdal equation for a spin-$s$ boson,
\be \label{Fronsdal0} \!\!\!\!\!\!\! \cF_{\m_1 \dots \m_s} \, \eq \,
\Box \vf_{\m_1 \dots  \m_s}  -  (\pr_{\m_1} \pr^{\, \rho} \vf_{\rho
\m_2 \dots \m_s} + \ldots )  + (\pr_{\m_1}  \pr_{\m_2}
\vf^{\,\rho}_{\, \, \rho \m_3 \dots  \m_s}
 +  \ldots)  = 0,
\ee
where terms completing the symmetrizations are left implicit, the
gauge field $\vf_{\, \m_1 \, \dots \, \m_s}$ is a totally symmetric
rank-$s$ tensor subject to the condition that its double trace
 $ \vf^{\rho\, \sigma}_{\hspace{12pt} \rho \, \sigma \,
\m_5  \dots  \m_s}$ \emph{vanish identically}, while the parameter
$\L_{\mu_1 \ldots \mu_{s-1}}$ entering the gauge transformation $ \d
\, \vf_{\, \m_1 \, \dots \, \m_s} \, = \, \pr_{\, \m_1} \, \L_{\,
\m_2 \, \dots \, \m_s} \, + \, \ldots $, under which
eq.~(\ref{Fronsdal0}) is invariant, is a \emph{traceless} rank-$(s -
1)$ tensor. Similar conditions are to be met for fermions, whose
equations, for the corresponding cases of symmetric rank-$n$
spinor-tensors $\ps_{\, \m_1 \, \dots \, \m_n}$, are
\be \label{fangfronsdaleq} \cS_{\, \m_1 \, \dots \, \m_n} \, \eq \,
\dsll \, \ps_{\, \m_1 \, \dots \, \m_n} \, - \, (\pr_{\m_1}\,
\psisl_{\, \m_2 \, \dots \, \m_{n}} \, + \, \ldots) \, = \, 0\, .
\ee
In the fermionic case the \emph{triple $\g$-trace} of the gauge
field is required to vanish, $ \g^{\, \a} \, \g^{\, \b} \, \g^{\,
\g} \, \ps_{\, \a \b \g \m_4 \dots \, \m_n} \, \equiv 0$, while the
gauge parameter $\e_{\mu_1 \ldots \mu_{n-1}}$, entering the
transformation law $ \d \, \ps_{\, \m_1 \, \dots \, \m_n} \, = \,
\pr_{\, \m_1} \, \e_{\, \m_2 \, \dots \, \m_n} \, + \, \ldots \, , $
is a symmetric rank-$(n - 1)$ spinor-tensor whose \emph{$\g$-trace}
is required to vanish, $ \g^{\, \r} \, \e_{\, \r \, \m_2 \, \dots \,
\m_{n-1}} \, \eq \, 0\, . $ This class of higher-spin gauge fields,
while not exhaustive in more than four dimensions, is nonetheless a
convenient ground for arriving at general results, that can
typically be extended to cases of mixed symmetry without major
difficulties. Here, following a common practice,  in arbitrary
dimensions we shall often call loosely ``spin'' the rank $s$ of the
bosonic gauge fields, or the rank $n$ of the fermionic fields
augmented by $\frac{1}{2}$.

These trace and $\gamma-$trace constraints, that as we have stressed
are somehow a legacy of similar conditions on the massive fields of
\cite{fierzpauli}, appear somewhat unpleasing, and thus we were
motivated to reformulate the theory in such a way as to bypass them.
Actually, a positive answer to this query had already been given by
Pashnev, Tsulaia, Buchbinder and others in \cite{pt,bp}, where the
free dynamics of \emph{individual and unconstrained} higher-spin
gauge fields was first described via BRST techniques \cite{BRST}.
Their construction, however, requires many additional fields, their
total number increasing linearly with the spin. This is to be
contrasted with the approach resulting from our investigations
\cite{fs3,dario}, that is being reviewed here, that removes the
trace constraints from the Fronsdal and Fang-Fronsdal Lagrangians
for all spins via \emph{at most a pair of additional fields}.

A convenient starting point for our constructions are the
Lagrangians related to eqs.~(\ref{Fronsdal0}) and
(\ref{fangfronsdaleq}), here written for \emph{unconstrained} fields
that transform with \emph{unconstrained} gauge parameters. Whereas
under these conditions the Lagrangians \emph{are not} gauge
invariant, in Section \ref{lagrangians} we shall see that in the
bosonic case a rank-$(s-3)$ compensator field $\a_{\, \m_1 \dots
\m_{s-3}}$ transforming as $ \d \, \a_{\, \m_1 \dots \m_{s-3}} =
\L^{\r}_{\hspace{6pt} \r \, \m_1 \dots \m_{s-3}}$, suffices to
eliminate from the variation all terms involving the trace of the
gauge parameter. Interestingly, the remainder is then proportional
to the \emph{gauge invariant combination}
\be \vf^{\, \r\, \s}_{\hspace{12pt} \r \, \s \, \m_5 \, \dots \,
\m_s} \, - \, 4 \, \prd \a_{\, \m_5\dots \m_s} \, \, - \,
(\pr_{\m_5} \, \a^{\, \r}_{\hspace{6pt} \r \m_6 \dots \m_s} \, + \,
\ldots \ ) \, , \ee
and as a result once this last constraint is enforced by a Lagrange
multiplier $\b$ one ends up with \emph{fully gauge invariant}
Lagrangians, whose equations of motion, as reviewed in Section
\ref{equations}, propagate indeed the correct polarizations. Similar
results hold for fermionic fields as well: in all cases (including
actually the BRST formulations of \cite{pt,bp}, as shown in
\cite{st}), the field equations can be reduced to the local
non-Lagrangian compensator equations of \cite{fs2,st}, that in the
bosonic case read simply
\bea && \cF_{\mu_1 \ldots \mu_s} \, = \, 3\, \partial_{\mu_1}
\partial_{\mu_2} \partial_{\mu_3 }\, \alpha_{\mu_4 \ldots \mu_s} + \ldots \, , \nonumber \\
&& {\vf^{\rho\sigma}}_{\rho\sigma \mu_5 \ldots \mu_s} \, = \, 4\,
\partial \cdot \alpha_{\mu_5 \ldots \mu_s} \, +\, (
\partial_{\mu_5} {\alpha^{\rho}}_{\rho \mu_6 \ldots \mu_s} \, + \, \ldots \ ) \, , \label{compens} \eea
where $\alpha$ denotes the rank-$(s-3)$ compensator defined above.

In Section \ref{nonlocal} we shall review how the additional fields
can be eliminated without reintroducing any trace constraints: the
end result will be a \emph{non-local} theory along the lines of
\cite{fs1,fs2}. For instance, for spin three the non-local equations
one obtains are
\be \!\!\!\! \cF_{\m_1\m_2\m_3} \ - \ \fr{1}{3 \, \Box} \, \left(\,
\pr_{\m_1} \, \pr_{\m_2} \, {\cF^{\, \r}}_{\r \m_3} \, + \,
\pr_{\m_2} \, \pr_{\m_3} \, {\cF^{\, \r}}_{\r \m_1} \ + \,
\pr_{\m_3} \, \pr_{\m_1} \, {\cF^{\, \r}}_{\r \m_2} \, \right) \, =
\, 0 \, , \label{nlspin3} \ee
where $\cF$ is defined in (\ref{Fronsdal0}). The dynamical content
of this theory is again the usual one of the Fronsdal formulation,
as we shall see following \cite{fs1,fs2,dario}.

Given the relative simplicity of the local formulation, the actual
interest in the non-local equations is mostly driven by conceptual
issues. Their most appealing feature is indeed a direct link to the
\emph{geometry} of these higher-spin fields, since the non-local
equations obtained eliminating the additional fields confer a
dynamical meaning to the higher-spin curvatures introduced in
\cite{dewfr} by de Wit and Freedman, that do not play a direct role
in the Fronsdal formulation. As will be explained in Section
\ref{geometry}, these curvatures involve as many derivatives as the
spin of the basic fields, and for this simple reason they can only
enter two-derivative equations provided they are accompanied by
inverse powers of the d'Alembertian operator, but all resulting
non-localities can be consistently eliminated by a partial
gauge-fixing procedure involving the trace of the gauge parameter.
The end result of this procedure was anticipated in \cite{fs1}, and
has the suggestive feature of being a direct generalization of the
geometric equations known for the more familiar spin-one and
spin-two cases. Thus, if $\cR_{\m_1 \m_2 \m_3, \, \r_1 \r_2 \r_3}$
is the curvature for the spin-three case, fully symmetric within the
two sets of indices, it is possible to show that the non-local
equation (\ref{nlspin3}) can be written in the form
\be \fr{1}{\Box} \ \h^{\, \r_1 \r_2} \ \pr^\s \, \cR_{\, \s \r_1
\r_2, \, \m_1 \m_2 \m_3} \ = \ 0 \, , \label{nlspin4} \ee
where $\h^{\, \r_1 \r_2}$ denotes the flat Minkowski metric, a
direct generalization of the spin-one Maxwell equation. In a similar
fashion, for spin four, the first non-trivial even-spin case, the
non-local geometric equations take the form \cite{fs1,fs2}
\be \fr{1}{\Box} \ \h^{\, \r_1 \r_2} \ \h^{\, \r_3 \r_4}\ \cR_{\,
\r_1 \r_2 \r_3 \r_4, \, \m_1 \m_2 \m_3 \m_4} \ = \ 0 \, , \ee
in clear analogy with the linearized spin-two Einstein equation. It
should be appreciated that the spin-one and spin-two cases are the
only ones for which these geometric equations are local.

The need for a better understanding of the relation between
Higher-Spin Gauge Theory and String Theory is a serious motivation
behind the interest in the former that is building up nowadays, and
some elementary related issues will be reviewed in Section
\ref{string}, where free equations emerging from String Field Theory
\cite{sft} in the low-tension limit will be compared with the
compensator equations (\ref{compens}). In that limit, as was known
for a long time, the structure of the String equations simplifies to
systems of three fields usually called ``triplets"
\cite{oldtriplet,fs2,bonelli} (or their fermionic analogues of
\cite{fs2}, or generalizations thereof \cite{oldtriplet,st}). In all
cases, the resulting fields are not subject to any trace constraints
and display a correspondingly unconstrained gauge symmetry. A
consistent truncation of the propagating modes then suffices to
recover the local compensator equations (\ref{compens}) for spin-$s$
bosons and corresponding ones for spin-($n+\12$) fermions. This
result and its generalizations to the mixed-symmetry case \cite{st}
can be taken as a hint that the unconstrained formulation should be
of help in unraveling the precise relationship between higher-spin
systems and String theory at the interacting level. A precise
statement to this effect, however, would still require further
investigation, in particular since at present the only window on
consistent higher-spin interactions is provided by the Vasiliev
construction \cite{vas1,vas2,vasrev}, which is strictly an on-shell
formulation. We shall therefore conclude in Section
\ref{vasilievtrace} with some cursory remarks on the role of the
unconstrained gauge symmetry in this construction, following
\cite{sss}. The key result will be that the extended symmetry can
indeed be accommodated in the free Vasiliev equations based on
vector oscillators \cite{vas2}, that are actually closer in spirit
to an off-shell form than their original version of \cite{vas1}
based on spinor oscillators. Interestingly, the compensators $\a$ do
not appear explicitly in the equations. Rather, they emerge as exact
forms of a generalized cohomological problem first discussed by
Dubois-Violette and Henneaux \cite{dubh}.

In order to simplify the discussion, in the following we shall
resort to a concise notation that we found quite useful in our
previous works. Thus, primes (or bracketed suffixes) will denote
traces, while all indices carried by the symmetric tensors
$\vf_{\mu_1 \dots \mu_s}$ and $\L_{\mu_1 \dots \mu_{s-1}}$, by the
metric tensor $\eta_{\mu\nu}$ or by derivatives will be left
implicit. In addition, all terms will be understood as totally
symmetrized, so that for instance $\pr \, \vf$ will stand for
$\partial_{\mu_1} \vf_{\mu_2 \ldots \mu_{s+1}} \, + \, \ldots$. A
few rules needed to take full advantage of this notation are:
\begin{eqnarray}
  \left( \pr^{\, p} \, \vf  \right)^{\, \pe} & = & \ \Box \,
  \pr^{\, p-2} \, \vf \ + \, 2 \, \pr^{\, p-1} \,  \prd \vf \ + \, \pr^{\, p} \,
\vf^{\, \pe} \,  , \nonumber \\
 \partial^{\, p} \, \partial^{\, q} & = & \ {p+q \choose p} % \binom{p+q}{p} \ \
\partial^{\, p+q} \ ,
\nonumber \\
 \partial \cdot  \left( \partial^{\, p} \ \vf \right) \ & = & \ \Box \
\partial^{\, p-1} \ \vf \ + \
\partial^{\, p} \ \partial \cdot \vf \ ,  \\
 \partial \cdot  \eta^{\, k} & = & \ \partial \, \eta^{\, k-1} \ , \nonumber\\
 \left( \eta^k \, \vf  \,  \right)^{\, \prime} & = & \ \left[ \, D
\, + \, 2\, (s+k-1) \,  \right]\, \eta^{\, k-1} \, \vf \ + \ \eta^k
\, \vf^{\, \prime} \, \nonumber . \label{etak}
\end{eqnarray}

For brevity, in the following Sections our discussion will be
restricted to bosonic fields, although all results presented extend
to fermionic fields, as discussed in
\cite{fs1,fs2,st,fs3,dario}\footnote[7]{The web site
http://www.ulb.ac.be/sciences/ptm/pmif/Solvay1proc.pdf contains the
Proceedings of the First Solvay Workshop on Higher-Spin Gauge
Fields, held in Brussels on May 12-14 2004, with a number of
contributions, including \cite{bd,fh,bcs,bkiv,sss} that are more
closely related to this work, and many additional references to the
original literature.}.
%%%%%%%%%%%%%%%%%%%%%%%%%%%%%%%%%%%%%%%%%%%%%%%%%%%%%%%%%%%%%%%%%%%%%

%%%%%%%%%%%%%%%%%%%%%%%%%%%%%%%%%%%%%%%%%%%%%%%%%%%%%%%%%%%%%%%%%%%%%
\section{``Minimal'' unconstrained Lagrangians for higher-spin fields}\label{lagrangians}

%%%%%%%%%%%%%%%%%%%%%%%%%%%%%%%%%%%%%%%%%%%%%%%%%%%%%%%%%%%%%%%%%%%%%

Let us consider the  Fronsdal Lagrangian \cite{fronsdal}
\be \cL_0 \, = \, \12 \ \vf \left(\ \cF \, - \, \12 \ \h \ \cF^{\,
\pe} \right) \, , \ee
written for an \emph{unconstrained} gauge field. Its
\emph{unconstrained} variation does not vanish, but
\begin{eqnarray} \label{var0}
\d \, \cL_0 \, = \,& \, \L^{\, \pe} \, \3s \, \left\{ \fr{3}{4} \
\prd {\cal{F}}^{\,
                              \pe} - \, \fr{3}{2} \ \prd \prd \prd \vf \, +\, \fr{9}{4} \, \Box \,
                              \prd \vf^{\, \pe} \, \right\} \nonumber \\
                           & - \,
                              \, 9 \ \4s \, \prd \prd \vf^{\, \pe}\, \prd \L^{\, \pe} \,
                              +\, \fr{15}{2} \ \5s \prd \prd \prd \vf^{\, \pe} \, \L^{\, \pe \pe} \\
                           &   - 3\ \4s \, \vf^{\, \pe \pe} \, \prd \prd \prd \L \, .  \nonumber
\end{eqnarray}
It is possible to add new terms involving the gauge field $\vf$ and
a spin-$(s-3)$ ``compensator'' field $\alpha$ in such a way that all
contributions involving the trace $\L^{ \, \pe}$ of the gauge
parameter disappear in the resulting variation \cite{fs3}. This
procedure does not eliminate all unwanted terms. Rather, it
generates a remainder proportional to the triple divergence of $\L$,
\be
\d \left\{ \cL_0 \, + \, \cL_1 \, +\, \cL_{2} \right\} \, = \,
 - \, 3 \ \{\vf^{\, \pe \pe} -\, 4 \, \prd
                     \a - \, \pr \, \a^{\, \pe} \} \, \prd \prd \prd \L \,
                     ,
\ee
that however is proportional to the \emph{fully gauge invariant}
combination
\be
 \vf^{\, \pe \pe} -\, 4 \, \prd \a \, - \, \pr \, \a^{\,
\pe} \, . \label{seccomp} \ee
This analysis rests on the Bianchi identity for the Fronsdal
operator,
\be \prd \cF \, - \, \12 \ \pr \, \cF^{\, \pe} \, = \, - \,
\fr{3}{2} \ \pr^{\, 3}\ \vf^{\, \pe \pe}\, , \label{bianchif} \ee
that contains a classical ``anomaly'' for spin $s>3$.

As we have recalled in the Introduction, the combination in
(\ref{seccomp}) plays a role in the compensator equations
(\ref{compens}) of \cite{fs2,st}. One can finally eliminate this
variation introducing another field, $\b$, that actually behaves as
a Lagrange multiplier for second of (\ref{compens}) and transforms
according to
\be \d \b \,= \, \prd \prd \prd \L \, , \ee
proportionally to the triple divergence of the gauge parameter. The
resulting Lagrangian for an \emph{unconstrained} spin-$s$ boson
\cite{fs3},
\begin{eqnarray}  \label{lagrange}
\cL= & \, \hspace{.1cm} \12 \, \vf  \, \left(\ {\cal F}  \, - \, \12
            \h  \, \cF^{\, \pe} \right) -  {s \choose 3} \a
           \left\{ \fr{3}{4} \, \prd {\cal F}^{\, \pe} \, - \, \fr{3}{2}\,
           \prd \prd \prd \vf + \fr{9}{4} \, \Box \, \prd \vf^{\, \pe} \right\}  \nonumber \\
        &+ \, 9 \, {s \choose 4} \prd \a  \, \prd \prd \vf^{\, \pe} -
           \fr{15}{2}  {s \choose 5}  \a^{\, \pe}\,
           \prd \prd \prd \vf^{\, \pe} + \fr{9}{4} \3s \a \, \Box^2 \, \a  \nonumber\\
        &- 27 \4s \prd \a \, \Box \, \prd \a  +  45 \5s  (\prd \prd \a)^2
          + \fr{45}{2}  \5s  \prd \prd \a  \, \Box \, \a^{\, \pe}  \nonumber\\
        & -  45 \6s  \prd \prd \prd \a  \, \prd \a^{\, \pe} +3 \4s
            \b  \left(\vf^{\, \pe \pe} -  4 \prd \a  -
           \pr \, \a^{\, \pe}\right) \ ,
\end{eqnarray}
is then invariant under the \emph{unconstrained} gauge
transformations
\begin{eqnarray}
& \d \, \vf\, = \, \pr \, \L \, , \nonumber\\
& \d \, \a \, = \, \L^{\, \pe} \, , \\
& \d \, \b \, = \, \prd \prd \prd \L \, .\nonumber
\end{eqnarray}

There is an alternative, simpler, way to build these unconstrained
Lagrangians. The starting point is in this case
\be \cL_0 \, = \, \12  \ \vf \, \left(\cA \, - \,  \12 \, \h\,
\cA^{\, \pe} \right)\, , \ee
where the \emph{fully} gauge invariant tensor $\cA$, defined as
\be \cA \, = \cF \, - \, 3 \, \pr^{\, 3} \, \a  \,  , \ee
enters the first of eqs.~(\ref{compens}). As was the case for the
Fronsdal operator $\cF$ (see eq.~(\ref{bianchif})), its Bianchi
identity contains a violation related to the double trace of the
gauge field $\vf$, that does not allow one to regard $\cA - \eta/2
\cA^\prime$ as a proper linearized Einstein-like tensor for higher
spins. The key difference with respect to the Fronsdal operator,
however, is that now the fully gauge invariant combination
(\ref{seccomp}) is involved, so that
\be \prd \cA \, - \, \12 \, \pr \, \cA^{\, \pe} \, = \, - \,
\fr{3}{2} \ \pr^{\, 3}(\vf^{\, \pe \pe}\, -\, 4\, \prd \a \, - \,
\pr \, \a^{\, \pe}). \ee
The trial Lagrangian $\cL_0$ has therefore a simple variation, that
can be nicely expressed in terms of a pair of \emph{gauge-invariant
quantities}:
\be \d \cL_0 \, = \, \fr{3}{4} {\small s\choose 3} \L^{\, \pe} \prd
\cA^{\, \pe} + 3 {s\choose 4} \prd \prd \prd \L \, (\vf^{\, \pe
\pe}-4\prd \a - \pr \a^{\, \pe})\, . \ee
In the same spirit as above, but in a simpler fashion, it is now
possible to compensate this remainder introducing the fields $\a$
and $\b$, and the complete Lagrangian is then simply
\be \!\!\!\!\!\!\!\!\!\! \cL = \, \12 \, \vf \, \left(\cA \, - \,
\12 \, \h \, \cA^{\, \pe} \right) - \fr{3\; \a}{4} {\small s \choose
3} \ \prd \cA^{\, \pe} - 3 \; \b \, {s\choose 4} (\vf^{\, \pe
\pe}-4\prd \a - \pr \a^{\, \pe}) \, . \ee

This alternative way of presenting (\ref{lagrange}) has the virtue
of allowing handy generalizations to the cases of (A)dS backgrounds,
that will presented elsewhere \cite{fms}. Similar results that hold
for fermion fields are not discussed here for brevity, but can be
found in \cite{fs3,dario,fms}.

%%%%%%%%%%%%%%%%%%%%%%%%%%%%%%%%%%%%%%%%%%%%%%%%%%%%%%%%%%%%%%%%%%%%%

%%%%%%%%%%%%%%%%%%%%%%%%%%%%%%%%%%%%%%%%%%%%%%%%%%%%%%%%%%%%%%%%%%%%%

\section{Structure and meaning of the field equations}\label{equations}

%%%%%%%%%%%%%%%%%%%%%%%%%%%%%%%%%%%%%%%%%%%%%%%%%%%%%%%%%%%%%%%%%%%%%

 In the previous Section we have reviewed how invariant bosonic
Lagrangians can be built introducing at most two additional fields.
Differently from the Fronsdal fields, the gauge fields $\vf$
involved in this description are not subject to any trace
constraints, and in order to check the consistency of the result one
would thus like to display the dynamical content of
eq.~(\ref{lagrange}). To this end let us consider the field
equations, and let us show that they can be reduced to the Fronsdal
equations (\ref{Fronsdal0}) by a partial gauge fixing.

The equations following from the Lagrangian (\ref{lagrange}) for the
fields $\vf$ , $\b$ and $\a$ are of the form
\begin{eqnarray} \label{abc1}
& \cA\, - \, \12 \, \h\, \cB \, + \, \h^2 \, \cC \, = 0\, , \nonumber \\
&\vf^{\, \pe \pe} \, - \, 4\, \prd \a \, - \, \pr \, \a^{\, \pe} \, = 0 \, ,  \\
& \cG_{\vf, \b}(\a) \, = \, 0 \, , \nonumber
\end{eqnarray}
with
\begin{eqnarray}
\cA \, & \eq \, \cF \, - \, 3 \, \pr^{\, 3} \, \a \, ,
\end{eqnarray}
where the explicit expressions for $\cB$, $\cC$ and $\cG$, not
needed for the present discussion, can be found in \cite{fs3}. It
should be noted that, when $\b$ is on-shell, the first of (\ref
{abc1}) reduces to
\be \label{abc} \cA\, - \, \12 \, \h\, \cA^{\, \pe} \, + \, \h^2 \,
\cC \, = \, 0 \, , \ee
while the double trace of $\cA$ vanishes \emph{identically}. A
straightforward analysis \cite{fs3,dario} then shows that the
equation for $\vf$ implies the independent relations $\cC = 0$,
$\cA^{\, \pe} = 0$, and finally the first of (\ref{compens}),
\be \cA \, \eq \cF - \, 3 \, \pr^{\, 3} \, \a \, = \, 0 \, . \ee
Once this form is reached, the Fronsdal equation (\ref{Fronsdal0})
can be directly recovered after a partial gauge fixing, making use
of the trace $\L^\pe$ of the gauge parameter, that can shift away
the field $\a$, while the second of (\ref{abc1}) removes at the same
time the double trace of the gauge field $\vf$.

This discussion completes the proof that the Fronsdal theory is
gauge equivalent to the unconstrained formulation. Nonetheless, the
unconstrained Lagrangians possess a wider gauge symmetry, and one
may wonder whether this could play a role in the non-linear
deformations of the theory.

The first hint that the wider gauge symmetry can imply some relevant
differences with the Fronsdal theory as soon as one moves away from
the free case can be appreciated comparing the properties of
\emph{external currents} coupled to the physical fields in the two
approaches. Let us indeed introduce the coupling to an external
source $\cJ$ in the standard fashion, adding to the Lagrangian
(\ref{lagrange}) a $\vf \cdot \cJ$ term. The equations of motion
become in this case
\begin{eqnarray} \label{abcj}
& \cA\, - \, \12 \, \h\, \cB \, + \, \h^2 \, \cC \, = \cJ\, , \nonumber\\
&\vf^{\, \pe \pe} \, - \, 4\, \prd \a \, - \, \pr \, \a^{\, \pe} \, = 0 \, ,  \\
& \cG_{\vf, \b}(\a) \, = \, 0 \, , \nonumber
\end{eqnarray}
and it is possible to show that the divergence of the left-hand side
of the first equation is proportional to the operator $\cG$ defining
the field equation of $\a$ \cite{fs3,dario}. Explicitly, the
divergence of the first equation in (\ref{abcj}) gives $ \fr{\h}{4}
\, \cG_{\vf, \b}(\a) \, =\, \prd \cJ \, , $ and therefore, because
of the third of (\ref{abcj}), \emph{the source must be
divergence-free on-shell}, as expected from Noether's theorem. This
result shows that the properties of the unconstrained system are
consistent with physical expectations for the most natural
interactions in a Field Theory, and marks a sizeable difference with
the Fronsdal scheme. In this last case, in fact, the corresponding
equations are
\be {\cal F} \ - \ \frac{1}{2} \ \h \ {\cal F}^{\, \pe}  \, = \, \cJ
\, , \ee
and consequently their divergence yields $
 -\, \12 \, \h \, \prd \cF^{\, \pe} \, = \, \prd \cJ \, .
$ Hence, while for $s\, = \,1, 2$ the usual condition $\prd \cJ \, =
\, 0$ is recovered, in Fronsdal's theory from spin $s\, = \, 3$
onwards consistency only implies the weaker condition that \emph{the
traceless part of the divergence} vanish. Although sufficient to
guarantee that only physical polarizations contribute to the
exchange of quanta between sources \cite{fronsdal}, this does not
completely fit the conditions for a Noether current. An equivalent
argument can be given for fermions, as in \cite{fs3,dario}.

%%%%%%%%%%%%%%%%%%%%%%%%%%%%%%%%%%%%%%%%%%%%%%%%%%%%%%%%%%%%%%%%%%%%%

%%%%%%%%%%%%%%%%%%%%%%%%%%%%%%%%%%%%%%%%%%%%%%%%%%%%%%%%%%%%%%%%%%%%%

\section{Removing the additional fields: non-local formulation}\label{nonlocal}

%%%%%%%%%%%%%%%%%%%%%%%%%%%%%%%%%%%%%%%%%%%%%%%%%%%%%%%%%%%%%%%%%%%%%%

In this Section we would like to show that it is possible to
eliminate the auxiliary fields from the unconstrained theory to
recover a formulation which is closer in spirit to the familiar
cases of spin one or two \cite{fs2,dario}. The resulting dynamics is
not local, but it is again possible to reduce it to the local
Fronsdal form by a partial gauge fixing involving the trace of the
gauge parameter.

The idea is to look for minimal combinations of the $\cF$ operator
with its traces and divergences, not involving $\a$ and such that
the resulting expression vanishes identically when the first of
(\ref{compens}) is satisfied. In the spin-$3$ case one can find in
this fashion two independent, fully gauge invariant, unconstrained
equations of this type,
\begin{eqnarray} \label{nonloc3}
&\cF \, -\, \fr{1}{3}\,\fr{\pr^{ \, 2}}{\Box}\,\cF^{\, \pe}\,\hspace{6pt} =\, 0 \, ,\\
&\cF \, -\, \fr{\pr^{\, 3}}{\Box^2}\,\prd \cF^{\, \pe} \,
\hspace{2pt} = \, 0 \, .
\end{eqnarray}
The second can be obtained combining the first with its trace, but
is anyway quite interesting, since it is of the form
\be \label{nonlocH} \cF \, = \, 3 \, \pr^3 \, \cH(\vf) \, , \ee
where $\cH(\vf)\, = \, \fr{1}{3\, \Box^2}\,\prd \cF^{\, \pe} \,$ is
such that $\d \, \cH(\vf) \, = \, \L^{\, \pe}$. In other words, the
non-local construct $\cH$ plays the role of the compensator $\a$!

We have thus seen that in the spin-3 case it is possible to
eliminate $\a$ from the compensator equation arriving at a fully
gauge invariant equation for the single field $\vf$, without
reintroducing any trace constraints. The irreducible equation is not
local, as anticipated, but can be cast into the particularly useful
form (\ref{nonlocH}), in which it is manifest that using only of the
trace $\L^\pe$ of the gauge parameter \emph{all non localities can
be removed}, thus recovering the Fronsdal form $\cF\, = \, 0$.

For bosons of arbitrary spin $s$ one can  reproduce the scheme just
described for spin $3$: starting from the first of (\ref{compens}),
it is possible to find a linear combination of constructs built from
$\cF$ that vanishes identically and does not involve $\a$, without
reintroducing any trace constraints \cite{fs1,fs2}. To this end, let
us define $ \cF^{(1)} \, \eq \, \cF$ and let us consider the
sequence of kinetic operators
\be \label{kinetic} {\cal F}^{(n+1)} \, = \, \cF^{(n)} \, + \,
\fr{1}{(n+1) (2 n + 1)} \, \fr{\pr^{\, 2}}{\Box} \, \cF^{(n)\, \pe}
\, - \, \fr{1}{n+1} \, \fr{\pr}{\Box} \ \prd \cF^{(n)} \ . \ee
They possess the crucial property that we need, since
eqs.~(\ref{compens}) and (\ref{kinetic}) imply that
\be \d \, \cF^{(n)} \, = \, \left( 2 n + 1 \right) \, \fr{\pr^{\; 2
n + 1}}{\Box^{\,n-1}} \, \a^{[n-1]} \, . \label{fngauge} \ee
Hence, for any spin $s$, if $p$ denotes the integer part of
$\{\fr{s-3}{2}\}$, the trace $\a^{[p+1]}$ of the compensator is
simply not available, and hence the non-local kinetic operator
$\cF^{(p+2)}$ satisfies the irreducible gauge-invariant equation
\be \label{kineticeq} \cF^{(p+2)} \, = \, 0 \, , \ee
where the compensator $\a$ has disappeared. Moreover, these kinetic
operators satisfy the modified ``Bianchi identities'' \cite{fs1,fs2}
\be \label{bianchin} \prd {\cal F}^{(n)} \, - \, \frac{1}{2n} \, \pr
\, \cF^{(n)\, \pe} \, = \, - \, \left( 1 + \frac{1}{2n}  \right) \,
\frac{\pr^{\; 2n+1}}{\Box^{\; n-1}} \, \vf^{[n+1]} \, . \ee
where for $n$ sufficiently large the classical ``anomaly'' also
disappears, so that they eventually suffice to define for all
spin-$s$ fields fully gauge invariant analogues of the Einstein
tensor,
\be \label{einsttens} {\cal G}^{(n)} \ = \ \sum_{p \leq n} \
\frac{(-1)^p}{2^p \ p\; ! \, \left( {n \atop p} \right)} \ \eta^{\,
p} \ {\cal F}^{(n)\, [p]} \, , \ee
that have vanishing divergences like their spin-2 counterpart. From
${\cal G}^{(n)}$ one can then build generalized Lagrangians that are
fully gauge invariant without any restrictions on gauge fields or
gauge parameters. Making use repeatedly of eq.~(\ref{bianchin}), it
is also possible to show that eqs.~(\ref{kineticeq}) can always be
reduced to the compensator-like form (\ref{nonlocH}) and then
finally gauge fixed to the local Fronsdal form \cite{dario}. Again,
similar results hold for fermion fields, but are not discussed here
for brevity and can be found in \cite{fs1,fs2,dario}.

%%%%%%%%%%%%%%%%%%%%%%%%%%%%%%%%%%%%%%%%%%%%%%%%%%%%%%%%%%%%%%%%%%%%%

%%%%%%%%%%%%%%%%%%%%%%%%%%%%%%%%%%%%%%%%%%%%%%%%%%%%%%%%%%%%%%%%%%%%%

\section{Higher-spin geometry from the non-local formulation}\label{geometry}

%%%%%%%%%%%%%%%%%%%%%%%%%%%%%%%%%%%%%%%%%%%%%%%%%%%%%%%%%%%%%%%%%%%%%%

The non-local formulation, although dynamically equivalent to the
local one, is clearly less manageable, both at the classical and at
the quantum level. So, what advantages can it possibly have? Briefly
stated, as we shall review in this Section, \emph{the non-local
formulation encodes a geometric description, for all higher spins,
along lines familiar from the two cases of spin-one and spin-two
fields}.

In order to motivate and explain this statement, our starting point
will be provided by the hierarchy of ``connections'' introduced by
de Wit and Freedman in \cite{dewfr}, that again can be nicely
motivated by a closer look at the first non-trivial case of a spin-3
boson. In analogy with the two lower-spin cases, the key idea is to
consider a linear combination of first derivatives of the basic
field $\vf_{\a \b \g}$ such that the resulting gauge transformation
be as simple as possible. The proper choice is
\be
 \G^{(1)}_{\r,\a\b\g} \, = \, \pr_{\r} \, \vf_{\a\b\g} - (\pr_{\a}\vf_{\r\b\g}+
                                                       \pr_{\b}\vf_{\r\a\g}+
                                                       \pr_{\g}\vf_{\a\b\r}) \ ,
\ee
since indeed
\be
 \d \G^{(1)}_{\r,\a\b\g} \, =\, -\, 2(\pr_{\a}\pr_{\b} \L_{\r\g} +
                                    \pr_{\a}\pr_{\g} \L_{\r\b} +
                                    \pr_{\b}\pr_{\g} \L_{\r\a}) \ ,
\ee
where for the sake of clarity all indices have been displayed. It is
possible to simplify further the gauge transformation taking a
combination of \emph{second} derivatives of $\vf$: the proper choice
is now
\be
 \G^{(2)}_{\r\s,\a\b\g}     = \pr_{\r} \Gamma^{(1)}_{\s,\a\b\g} -
                                   \12(\pr_{\a} \Gamma^{(1)}_{\s,\r\b\g}+ \dots )  \, ,
\ee
since the resulting variation is simply
\be
 \d \, \G^{(2)}_{\r\s,\a\b\g} = 3 \, \pr_{\a}\pr_{\b}\pr_{\g}\L_{\r\s} \, ,
\ee
that resembles more closely the behavior of the spin-2 Christoffel
connection or of the spin-1 gauge field. In this respect, for spin
$3$ this $\Gamma^{(2)}$, more than $\Gamma^{(1)}$, is what one would
properly call a ``connection", and indeed a sort of ``symmetric
curl" of $\Gamma^{(2)}$ builds a gauge invariant ``curvature" for
the spin $3$ field $\vf$:
\be
 \Gamma^{(3)}_{\, \r\s\t,\, \a\b\g}
 \eq  \cR^{(3)}_{\, \r\s\t,\, \a\b\g}\, = \, \pr_{\, \r} \Gamma^{(2)}_{\, \s\t,\, \a\b\g} -
                                   \fr{1}{3}(\pr_{\, \a} \Gamma^{(2)}_{\, \s\t,\, \r\b\g}+
                                   \dots )\, .
\ee

In order to deal with the general case, one can follow a strategy
that should be clear from the previous example, the key observation
being that in each $\G^{(m)}_{\, \r_1 \, \dots \, \r_m, \, \a_1 \,
\dots \, \a_s}$ the two groups of indices \emph{are not equivalent}.
The $\r$'s are ``special" indices, related to the introduction of
$m$ derivatives at the $m$-th step, and the combinations should be
chosen in such a way that in the gauge variation of $\G^{(m)}$ the
special indices end within the gauge parameter $\L_{\, \m_1 \, \dots
\, \m_{s-1}}$. If this is attained, in the gauge transformation of
$\G^{(s-1)}$ \emph{all} indices belonging to the gauge parameter
will be special ones, and consequently $\G^{(s)}$ (along with any
$\G^{(s+k)}$ for $k
>0$) will be necessarily gauge invariant, simply because there will
be no room in $\L$ to accommodate more special indices. In order to
present these results in a compact fashion, let us resort to a mixed
notation, in which the $``\pr"$ symbol is reserved for derivatives
with respect to ``special'' indices while $``\btd"$ is taken to
denote derivatives with respect to the remaining ones. Bearing in
mind that symmetrization only applies to pairs of indices within
each of the two sets, one could write the de Wit-Freedman symbols in
the form
\be \G^{(m)} \, = \, \sum_{k=0}^{m}\fr{(-1)^{k}}{ \left( {{m} \atop
{k}} \right) }\ \pr^{\, m-k}\btd^{\, k} \vf\, ,\ee
and the corresponding gauge transformations,
\be
 \d \, \G^{(m)} \, = \, (-1)^m \,
(m+1) \, \btd^{\, m+1} \,  \L \, , \ee
then display their direct correspondence with the kinetic operators
$\cF^{(n)}$ of the previous Section (see eq.~(\ref{fngauge}). In
particular, in this notation the gauge-invariant curvatures become
\be \label{curv} \cR^{(s)} \, = \, \sum_{k=0}^{s}\fr{(-1)^{k}}{
\left( {{s} \atop {k}} \right) }\ \pr^{\, s-k}\btd^{\, k} \vf \, .
\ee

These curvatures \emph{can not} be used directly to describe the
Fronsdal dynamics, simply because they are higher-derivative
objects, and this was the reason why the authors of \cite{dewfr}
rejected the possibility of describing higher-spin fields in an
\emph{unconstrained} fashion, returning at the end of their analysis
to the constrained Fronsdal formulation. And indeed, if one assumes
that $\L^{\, \pe} \eq 0$, the trace of each connection in the
hierarchy is also gauge invariant, and in particular the trace of
the $\G^{\, (2)}$ defines a gauge invariant second order operator,
that coincides with $\cF$ and in \cite{dewfr} was chosen to define
the field equation, thus justifying Fronsdal's result from a
different viewpoint.

Nonetheless, one can well try and describe via the curvatures
(\ref{curv}) an ``effective'' second-order dynamics, acting on them
with suitable inverse powers of the d'Alembertian operator in order
to recover the physical dimensions of a second-order operator. The
resulting \emph{geometric} equations \cite{fs1,fs2}
\begin{eqnarray}
\hspace{1.5cm} & \fr{1}{\Box^{n-1}} \, \cR^{[n]}{}_{\m_1 \cdots
\m_{2n}} \, = \, 0 \, ,   & (s\, = \, 2\, n) \, , \hspace{1.5cm}
\label{geomeven}\\
\hspace{1.5cm} & \fr{1}{\Box^{n}} \, \prd \cR^{[n-1]}{}_{\m_1 \cdots
\m_{2n-1}} \, = \, 0 \, ,
 \quad & (s\, = \, 2\, n \, + \, 1)\, ,\label{geomodd} \hspace{1.5cm}
\end{eqnarray}
are a nice way of encoding the results of the previous Section, once
the curvatures are \emph{identified} with suitably iterated
$\cF^{(n)}$ operators. These geometric equations are natural
generalizations of the well-known spin-1 and spin-2 cases, are of
course non-local on account of our preceding arguments, but as we
have reviewed in previous Sections the non local terms are harmless
gauge artifacts, as was the case for the non-local terms in the
equivalent kinetic operators (\ref{kinetic}). These equations were
generalized to the case of mixed symmetry in \cite{bb,dem}.

In conclusion, \emph{the curvatures of de Wit and Freedman acquire a
direct dynamical meaning in the non-local theory}, whose equivalence
to the local unconstrained one has been discussed in the previous
Section. This description embodies the linearized forms of the
non-linear Yang-Mills and Einstein equations, for spin $1$ and $2$
respectively, and leads naturally to speculate that a currently
unknown non-linear metric-like geometry (to be contrasted with
Vasiliev's frame-like geometry \cite{vas1,vas2,vasrev}) could
alternatively be taken to underlie higher-spin interactions, in such
a way as to give rise to eqs.~(\ref{geomeven}) and (\ref{geomodd})
in the linearized limit. Similar results hold for fermionic fields,
and are discussed in some detail in \cite{dario}.
%%%%%%%%%%%%%%%%%%%%%%%%%%%%%%%%%%%%%%%%%%%%%%%%%%%%%%%%%%%%%%%%%%%%%

%%%%%%%%%%%%%%%%%%%%%%%%%%%%%%%%%%%%%%%%%%%%%%%%%%%%%%%%%%%%%%%%%%%%%

\section{Higher-spin geometry and String Field Theory}\label{string}

%%%%%%%%%%%%%%%%%%%%%%%%%%%%%%%%%%%%%%%%%%%%%%%%%%%%%%%%%%%%%%%%%%%%%%

Higher-spin states are an intrinsic, inevitable part of the string
spectrum. Hence, it is natural to look for links between these
properties of higher-spin fields and what can be deduced from String
Field Theory in the linearized approximation.

In this Section we would thus like to review briefly how the
unconstrained formulation of free higher-spin gauge fields can be
related to the low-tension limit of free String Field Theory. We
shall depart slightly from \cite{fs2,st}, and follow \cite{dario}.
As shown in \cite{sft}, in all cases the free equations of String
Field Theory can be written in the form
\begin{equation}
{\cal Q} \ | \Phi \rangle \ = \ 0 \ , \label{eqstring}
\end{equation}
where ${\cal Q}$ is the BRST operator of the first-quantized string.
These systems display the chain of  gauge invariances
\begin{equation}
\delta | \Phi^{(n)} \rangle \ = \ \ {\cal Q} \, | \Phi^{(n+1)} \rangle \ ,
\label{gaugestring}
\end{equation}
with $\Phi^{(1)}$ the string gauge field and $\Phi^{(n)} \ ( n > 1)$
a corresponding chain of gauge parameters.

In the usual case of tensile strings, these equations describe
\emph{massive} higher-spin modes, with masses determined by the
string tension, but it is important to stress that \emph{neither}
(\ref{eqstring}) \emph{nor} (\ref{gaugestring}) \emph{involve trace
conditions}. Hence, it is reasonable to expect that in the
low-tension limit $\a^{\, \pe} \to \infty$ the resulting massless
dynamics should naturally relate to the unconstrained formulation
reviewed in the previous Sections, rather than to Fronsdal's
constrained equations.

Even for the open bosonic string, the complete analysis of the
spectrum would require to deal with mixed-symmetry tensors, as
discussed in \cite{st}. Here it will suffice to note that for
symmetric tensors (corresponding to states in the leading Regge
trajectory of the open bosonic string, generated by the lowest
string oscillators $\alpha_{-1}$) the limit $\a^{\, \pe} \to \infty$
yields \emph{triplet} systems of the type
\cite{oldtriplet,fs2,bonelli}
\begin{eqnarray}
&\Box \, \f \, = \, \pr \, C \, ,  \nonumber \\
&\pr \, \cdot \, \f \, - \, \pr \, D \, = \, C \, , \label{bosestring} \\
&\Box \, D \, = \, \pr \, \cdot \, C \ ,\nonumber
\end{eqnarray}
where $\f$, $C$ and $D$ are symmetric tensors of ranks $s$, $s-1$
and $s-2$ respectively. Notice that \emph{no trace constraints are
enforced on these fields}, while these equations are invariant under
the {\it unconstrained} gauge transformations
\begin{eqnarray}
& \d \, \f \, = \, \pr \, \L \, ,  \nonumber \\
& \d \, C \, = \, \Box \, \L \, ,   \\
& \d \, D \, = \, \prd \L \, .\nonumber
\end{eqnarray}

In order to establish contact with the unconstrained equations of
the preceding Sections, one can proceed as in \cite{fs1}, or
alternatively one can manipulate eq.~(\ref{bosestring}) in order to
reproduce the Fronsdal operator (\ref{Fronsdal0}), as in
\cite{dario}. In particular, making use of the second of
(\ref{bosestring}) to express the gradient of $\prd \f$ in terms of
$C$ and $D$ according to
\be \label{triplet1} \pr \, \prd \f \, = \, 2 \, \pr^{\, 2} \, D \,
+\, \pr  \, C \, , \ee
and taking a double gradient of the trace of the first, after
dividing by the d'Alembertian operator one can arrive at
\be \label{triplet2} \pr^{\, 2} \, \f^{\, \pe} \, = \, 2 \, \pr^{\,
2} \, D \, + \, 3 \, \fr{\pr^{\, 3}}{\Box} \, C^{\, \pe} \, . \ee
Putting together (\ref{triplet1}) and (\ref{triplet2}) with the
first of (\ref{bosestring}), the final result,
\be
 \cF \, = \, 3 \,  \fr{\pr^{\, 3}}{\Box} \, C^{\, \pe} \, ,
\ee
is clearly of the form (\ref{compens}), with the identification of
$\a$ with $\fr{1}{\Box} \, C^{\, \pe}$. And indeed, starting from
the third of the triplet equations one can also arrive at
\be \f^{\, \pe \pe} \, = \, 4 \, \fr{1}{\Box}\ \prd C^{\, \pe} \, +
\, \fr{\pr }{\Box} \ C^{\, \pe \pe} \, . \ee

Notice that in this fashion one has truncated all modes that are
responsible for the propagation of lower spins $(s-2)$, $(s-4)$,
$\ldots$ in the triplet, so that the end result is to exhibit a
direct link between the (truncated) triplet (\ref{bosestring}) and
the compensator equations (\ref{compens}). To reiterate, in the
low-tension limit the equations of String Field Theory for fully
symmetric tensors bear a direct relationship to the unconstrained
formalism for higher spins reviewed in Section 2 of this paper, and
hence to the non-local geometric equations reviewed in Section 3.
Similar results hold for mixed-symmetry tensors \cite{st}, and for
fermion fields \cite{fs2,st,dario}, but are not discussed here for
the sake of brevity.

%%%%%%%%%%%%%%%%%%%%%%%%%%%%%%%%%%%%%%%%%%%%%%%%%%%%%%%%%%%%%%%%%%%%%

%%%%%%%%%%%%%%%%%%%%%%%%%%%%%%%%%%%%%%%%%%%%%%%%%%%%%%%%%%%%%%%%%%%%%

\section{Unconstrained gauge symmetry and the Vasiliev equations}\label{vasilievtrace}

%%%%%%%%%%%%%%%%%%%%%%%%%%%%%%%%%%%%%%%%%%%%%%%%%%%%%%%%%%%%%%%%%%%%%%

At present not much is known in a systematic fashion about
higher-spin interactions, aside from the fact that they are bound to
involve infinitely many fields, but a remarkable example is
available. \emph{Consistent non-linear interactions} for an infinite
set of fully symmetric tensors can be elegantly encoded in the
Vasiliev equations \cite{vas1,vas2,vasrev}, a set of curvature
constraints relating a one-form master field $A$ containing the
higher-spin fields and a zero-form master field $\Phi$ subsuming the
content of corresponding Weyl curvatures (as well as an independent
scalar mode). These fields may also be matrix-valued, in analogy
with standard Chan-Paton \cite{chanpaton} constructions, which
suggests a natural link with the leading Regge trajectory of open
bosonic strings. The key idea behind these equations is to bypass
the standard $S$-matrix no-go theorems for higher-spin interactions
by introducing a cosmological term \cite{fradkvas}, that builds
naturally an expansion about dS or AdS backgrounds rather than
around flat space, and to generalize the frame formulation of
gravity to higher spins by properly extending the $SO(1,D)$
($SO(2,D-1)$) tangent-space Lorentz algebra of $(A)dS_D$ Einstein
gravity. Thus, while the latter can be realized via quadratic
expressions in oscillators, the Vasiliev formulation rests on the
use of arbitrary polynomials, that essentially define what are
commonly called ``higher-spin algebras''.

There are actually two distinct versions of the Vasiliev equations.
The original one \cite{vas1} applies to the four-dimensional case,
rests on the properties of Grassman-even oscillators $\xi_\alpha$
and $\xi_{\dot{\alpha}}$ that are spinors of the tangent-space
Lorentz group, and if truncated results in free equations that are
precisely in the constrained Fronsdal form, although they are partly
presented in spinor notation. On the other hand, a more recent
version \cite{vas2} rests on a different set of Grassman-even
oscillators, $Y^i_A$, that are simultaneously Lorentz vectors and
doublets of an internal $Sp(2,R)$. It is closer in spirit to an
off-shell form and, interestingly, can accommodate the extended
symmetry reviewed in this paper. The internal $Sp(2,R)$ symmetry
brings about some subtleties, requires suitable projections, and
allows two distinct options. The first, usually called ``weak
projection'', pursued by Vasiliev in \cite{vas2} and reviewed in
detail in \cite{vasrev}, treats in a similar fashion the one-form
$A$ and the zero-form $\Phi$, and at the free level reduces again to
the Fronsdal formulation. The second option, usually called ``strong
projection'' \cite{sss}, restricts \emph{only the zero form} $\Phi$
while letting the gauge fields adjust correspondingly.

Once trace constraints are not enforced on the gauge fields, it is
natural to expect that the compensators $\a$ should play a role, but
this happens in a rather amusing and surprising fashion. In the
following we shall try to give a flavor of this result, following
\cite{sss}. We should stress, however, that the prospects for the
``strong'' projection at the interacting level are not fully clear
at the moment, since it apparently brings about Weyl-ordering
divergences in non-linear terms, but on the other hand the off-shell
origin of the Vasiliev equations, where the unconstrained
formulation would be expected to play a key role, is also unclear.
The recent work in \cite{sss} contains a preliminary analysis of the
problem. Although with the ``strong'' projection divergences would
naively manifest themselves in non-linear interactions, one can not
exclude at the present time that \emph{divergent field
redefinitions} or other subtleties may eventually suffice to remove
them\footnote[7]{A.S. is grateful to C. Iazeolla and P. Sundell for
extensive discussions on this point.}. A proper discussion of the
Vasiliev equations and of all these issues, however, is beyond the
scope of this review, and therefore we shall content ourselves with
providing some hints.

The key fact is that, in building master fields with the vector
oscillators $Y^i_A$ that satisfy the commutation relations
\be \left[ Y^i_A , Y^j_B \right] \ = \ i\, \e^{ij} \ \eta_{AB} \quad
(A,B=1,...,D+1) \quad (i,j=1,2) \label{bosealg} \ee
one ends up in principle with Taylor coefficients, the ordinary
fields, that aside from filling the desired $SO(2,D-1)$ or $SO(1,D)$
representations generally transform under $Sp(2,R)$ as well.

The $Sp(2,R)$ internal symmetry, on the other hand, reflects a
redundancy, its origin being the need to dispose of dual sets of
``coordinates'' (the  $Y^1_A$) and ``momenta'' (the $Y^2_A$) in
order to write eq.~(\ref{bosealg}). Hence the need for a restriction
to fields that are $Sp(2,R)$ singlets, that can be effected by
suitable constraints on $A$ and $\Phi$, but in order to arrive at
dynamical equations one is to go further, removing also traces from
the Weyl tensors in $\Phi$, for a reason that we shall now try to
explain. To this end, following \cite{sss}, and abiding to a similar
notation, let us begin by recalling that, if all interactions are
neglected and the cosmological constant is subsequently contracted
away, the Vasiliev equations with a ``strong'' projection reduce to
the flat-space relations $(s \geq 2; k=0,\ldots,s-1)$
\cite{freeframe}
\be \partial_{[\mu} \, W^{(s-1,k)}_{\n],a(s-1),b(k)} \, + \,
W^{(s-1,k+1)}_{[\mu|,a(s-1),|\nu]b(k)}\nn\, = \, \delta_{k,s-1}\
C_{[\mu|a(s-1),|\nu]b(s-1)} \, .\label{eq:fr6}\ee

Here $W^{(s-1,k)}$ identifies a field transforming according to a
traceful two-row Young tableau, with $(s-1)$ boxes in the first row
and $k$ boxes in the second, in a convention where total row
symmetrization is manifest, while for instance $a(s-1)$ and $b(k)$
denote two sets of $(s-1)$ and $k$ fully symmetrized tangent-space
vector indices. Eqs.~(\ref{eq:fr6}) are invariant under the gauge
transformations
\begin{eqnarray}
&& \d W_{\m,a(s-1),b(k)}^{(s-1,k)} \ = \ - \partial_\m \,
\xi_{a(s-1),b(k)}^{(s-1,k)} \ + \
\xi_{a(s-1),b(k)\m}^{(s-1,k+1)}\quad (k=0,\ldots,s-2) \nonumber \\
&& \d W_{\m,a(s-1),b(s-1)}^{(s-1,s-1)} \ = \ - \partial_\m \,
\xi_{a(s-1),b(s-1)}^{(s-1,s-1)} \ ,
\end{eqnarray}
involving the sequence of gauge parameters
$\xi_{a(s-1),b(k)}^{(s-1,k)}$, for $k=0,\ldots,s-1$. It takes some
work to bring the compact-looking non-linear equations of
\cite{vas2},
\be \widehat F\ =\ {i\over 2} \ dZ^i\wedge dZ_i \ \widehat \Phi\star
\kappa\ ,\quad\quad \widehat D \, \widehat \Phi\ =\ 0\ ,
\label{master} \ee
to this form, since the former involve additional non-commutative
$Z$ coordinates to be disposed of via a perturbative expansion in
powers of $\Phi$. A nice discussion of how this can be achieved for
the simpler spinorial construction can be found in \cite{analysis}.
At any rate, eq.~(\ref{eq:fr6}) has a rather transparent meaning: it
equates Riemann-like tensors to Weyl-like ones. If the latter are
trace-free, the end result is then enforcing, in an indirect way
nicely compatible with gauge invariance, via the trace of
eq.~(\ref{eq:fr6}), higher-spin generalizations of the Einstein
equation ``$Ricci=0$''.

Insofar as the free limit is concerned, working with the ``strong''
projection of \cite{sss} amounts to allowing in eqs.~(\ref{eq:fr6})
one-forms $W$ subject to no trace conditions while restricting the
corresponding Weyl tensors in $C$ to be traceless. Let us consider
eq.~(\ref{eq:fr6}) for the two cases of $s=2$ and $s=3$, in order to
display explicitly what subtlety presents itself in the latter,
leading eventually to the emergence of the compensator $\a$.

The $s=2$ case is the familiar example of linearized Einstein
gravity in the frame formulation. Eqs.~(\ref{eq:fr6}) becomes in
this case a set of two conditions: the first, corresponding to
$k=0$, is the ``vielbein'' postulate, that relates the spin
connection $W_{\mu,a,b}^{(1,1)}$ to the vielbein
$W_{\mu,a}^{(1,0)}$, while the trace of the second, corresponding to
$k=1$, gives the Einstein equation. A partial gauge fixing of
$W_{\mu,a}^{(1,0)}$ to a symmetric $h_{\mu\nu}$ recovers the
linearized Einstein equation in the metric form, that is also the
$s=2$ case of the Fronsdal equation (\ref{Fronsdal0}). Notice that
the condition that this gauge choice be preserved correlates the two
gauge parameters $\xi_a$ and $\xi_{a,b}$, in the $(1,0)$ and
$(1,1)$, in such a way that
\be \xi_{a,b} \ = \ \frac{1}{2} \, \left( \partial_b \, \xi_a \ - \
\partial_a \, \xi_b \right) \, , \ee
which turns the gauge transformation of the vielbein
\be \delta \, W^{(1,0)}_{\mu a} \ = \ -\  \partial_\mu \xi_a \ + \
\xi_{a,\m} \ee
into a proper transformation for $h_{\m\n}$.

The novelty first shows up for spin $s=3$, since in this case the
iterative solution of the homogeneous constraints builds up at the
second step a second-order operator $W_\mu^{(2,2)}$, whose trace in
a pair of tangent-space indices has the right structure to be the
field equation. This is fully determined by its original gauge
transformation into the gradient of a parameter $\xi_{ab,cd}$
transforming in the $(2,2)$ of the tangent-space group, built from
the lowest gauge parameter $\xi_{ab}$ and two derivatives,
\be \xi_{ab,c}^c \ \sim \ \Box \, \xi_{ab} \ - \ \partial_a\,
\partial \cdot \xi_b \ - \ \partial_b\, \partial \cdot \xi_a \ + \
\partial_a\,
\partial_b \, \xi^\pe  \, ,
 \ee
so that
\be \eta^{cd} \ W^{(2,2)}_{ab,cd} \ = \  \Box \, \vf_{\m ab} \ - \
\partial_a\,
\partial \cdot \vf_{\m b} \ - \ \partial_b \, \partial \cdot \vf_{\m a} \ + \
\partial_a\,
\partial_b \, {\vf^\pe}_\m \, , \label{w22} \ee
after gauge fixing $W^{(2,0)}_{\m, ab}$ to a fully symmetric tensor
$\vf_{\m\n\r}$.

However, given the trace-free condition on the Weyl tensor $C$, the
$k=2$ condition implies the additional constraint
\be \partial_{[\mu} W^{(2,2)}_{\n],ab,c}{}^c\ =\ 0\, ,\ee
and therefore the term in eq.~(\ref{w22}) is to be \emph{a pure
gradient},
\be \Box
\vf_{ab\mu}-\partial_{a}\partial\cdot\vf_{b\mu}-\partial_{b}\partial\cdot\vf_{a\mu}
+\partial_a\partial_b\vf'_\mu\ \ = \ \partial_\mu\b_{ab} \ , \ee
but a curl of this implies an additional constraint on $\beta$:
\be \partial_{[\mu}(\partial\cdot \vf_{a]b}-\partial_{a]}\vf'_b)\ =\
\partial_{[\mu}\b_{b]a}\ ,\label{eq:s33}\ee
This constraint has an inhomogeneous solution,
\be \b_{ab} \ = \ (\partial\cdot \vf_{ab}-\partial_{a}\vf'_b -
\partial_{b}\vf'_a)\, ,\label{inhom}\ee
but also \emph{a homogenous solution}, and this is where the
compensator $\a$ actually sits, as an ``exact'' form in the sense of
Dubois-Violette and Henneaux \cite{dubh}.

Let us pause to see explicitly the subtlety. They key point is that,
if one tries to solve the condition
\be
\partial_{[\mu}\b_{a]b} \ = \ 0 \, , \label{constraintb}
\ee
the symmetry in $(a,b)$ \emph{does not allow} a single gradient.
This is simple to see since, letting
\be \b_{ab} \ = \ \partial_a \alpha_b + \partial_b \alpha_a \, , \ee
eq.~(\ref{constraintb}) implies indeed the additional constraint
\be \partial_b \partial_{[\mu} \alpha_{a]} \ = \ 0 \, , \ee
whose solution is
\be \alpha_a \ = \ \partial_a \alpha \, . \ee
This identifies the compensator $\a$, and putting these results
together one finally recovers the compensator equation
(\ref{compens}) for $s=3$. For higher spins one must similarly work
backwards from the highest constraint to the two-derivative
operator, \emph{retaining the exact terms} emerging at subsequent
steps and subjecting them to due constraints. The end result is in
the general the recovery of eq.~(\ref{compens}), with spin-$(s-3)$
compensators emerging from terms that are exact in the sense of
\cite{dubh}. This derivation resonates with a result found by
Bekaert and Boulanger \cite{bb} in their analysis of the link
between the compensator $\a$ and the hierarchy of Freedman-deWit
connections of \cite{dewfr}, but it is actually different since,
despite some unfortunate nomenclature in \cite{sss}, in the Vasiliev
construction one is working with generalized spin connections,
rather than with generalized Christoffel symbols.

%%%%%%%%%%%%%%%%%%%%%%%%%%%%%%%%%%%%%%%%%%%%%%%%%%%%%%%%%%%%%%%%%%%%
\section*{Acknowledgements}

We are grateful to J. Mourad, E. Sezgin and M. Tsulaia, and
especially to C. Iazeolla and P. Sundell, for extensive discussions
and/or collaboration on issues touched upon in this review. This
work was supported in part by INFN, by the EU contracts
MRTN-CT-2004-503369 and MRTN-CT-2004-512194, by the INTAS contract
03-51-6346 and by the NATO grant PST.CLG.978785. A.S. would like to
thank the CERN Ph-Th Department, where he is supported in part by
the Scientific Associates program, while D.F. would like to thank
MPI-AEI, where he is supported in part by an EU pre-doctoral
Fellowship. 

\end{document}